\documentclass{PoS}
                                                                               
\PoS{PoS(LAT2005)036}
                                                                                
\title{Asymptotic Scaling and Monte Carlo Data }
                                                                                
\ShortTitle{Asymptotic Scaling and Monte Carlo Data }

\author{\speaker{A. Trivini}\\
        University of Wales Swansea\\
        E-mail: \email{pyat@swan.ac.uk}}

\author{C. R. Allton\\
        University of Wales Swansea\\
        E-mail: \email{c.allton@swansea.ac.uk}}

\abstract{It is a generally known problem that the behaviour predicted from 
perturbation theory for asymptotically free theories like QCD,
i.e. asymptotic scaling, has not been observed in Monte Carlo simulations
when the series is expressed in terms of the bare coupling $g_0$.
This discrepancy has been explained in the past with the poor
convergence properties of the perturbative series in the  $g_0$.
An alternative point of view, called {\em Lattice-Distorted Perturbation Theory}
proposes that {\em lattice artifacts} due to
the finiteness of the lattice spacing $a$ cause the disagreement
between Monte Carlo data and perturbative scaling.
Following this alternative scenario, we fit recent quenched data from
different observables to fitting functions that include these cut-off
effects, confirming that the lattice data are well reproduced by  $g_0$-PT
with the simple addition of terms ${\cal O}(a^n)$. }

\FullConference{XXIIIrd International Symposium on Lattice Field Theory\\
                 25-30 July 2005\\
                 Trinity College, Dublin, Ireland}

\newcommand{\ia}{a^{-1}}                                                    
\newcommand{\x}{X_{n,\nu}}
\newcommand{\y}{Y_{n',\nu'}}
\newcommand{\be}{\begin{equation}}
\newcommand{\ee}{\end{equation}}

\begin{document}

                                                                               
\section{Asymptotic scaling}
\label{sec:a.scal.}

The so-called $\beta$-function quantifies the dependence of the coupling 
constant on the lattice spacing $a$: 

\begin{equation}
\beta(g^2) = -a \frac{dg^2}{da}.
\end{equation}
Integrating this expression one obtains a relation between $g$ and $a$ which is
the usual expression for the running of the coupling:

\begin{equation}
\ia (g^2) = \frac {\Lambda}{f_{PT}(g^2)}, 
\label{eq:ia}
\end{equation}
where
\begin{equation}
f_{PT}(g^2) = e^{- \frac{1}{2 b_0 g^2} } \;\; (b_0 g^2)^{-b_1 \over {2 b_0^2}}
 \;\; ( 1 + d_2 g^2 )
\label{eq:f_PT}
\end{equation}
is the 3-loop scaling function, $b_0, b_1$ and $d_2$
the usual 1, 2 and 3-loop coefficients, and $\Lambda$ is
a constant of integration. For any lattice prediction of QCD
to have physical relevance it should follow the asymptotic scaling 
according to Eqs.~(\ref{eq:ia}) and (\ref{eq:f_PT}), in the limit of the
bare coupling $g_0\rightarrow 0$.

Figure 1 is an example of the strong disagreement between Monte Carlo data 
and $g_0-$PT, where the data is taken from \cite{Necco}.\footnote
{The lattice spacing in this figure, $a_{r_c}$ is introduced in \cite{Necco}
and defined via the force, $F(r)$, in an analogous fashion to $r_0$, but with
$r_c^2 F_c(r_c) = 0.65$.}
We fit the same set of data in different 
renormalized schemes known in the literature;
the $g_{V}$ schemes were introduced in \cite{Lep&Mack},
the $g_E$ in \cite{Parisi},
and the $g_{E2}$ in \cite{Bali&Sch}.
We can see that the mismatch decreases using a 
renormalized coupling constant instead of the bare one \cite{Lep&Mack}.

However, a better reduction (see Fig. 1) is obtained with Lattice-Distorted PT 
\cite{Chris}, using $g_0$ as expansion parameter and including lattice
artifacts ${\cal O}(a^n)$ due to the systematic error in the
Monte Carlo data due to the finiteness of $a$.
The expression (\ref{eq:ia}) becomes:
\begin{equation}
\ia_L(g_0^2) = \frac{\Lambda_L}{f_{PT}(g_0^2)}
    \times  ( 1 + \sum_{n=1} c'_n(g_0^2) f_{PT}^{\;n}(g_0^2) ),
\label{eq:ia_LDPT}
\end{equation}
where $\Lambda_L$ is the scale parameter of lattice QCD.


\begin{figure}
\begin{center}
\includegraphics[angle=0,width=0.6\textwidth]{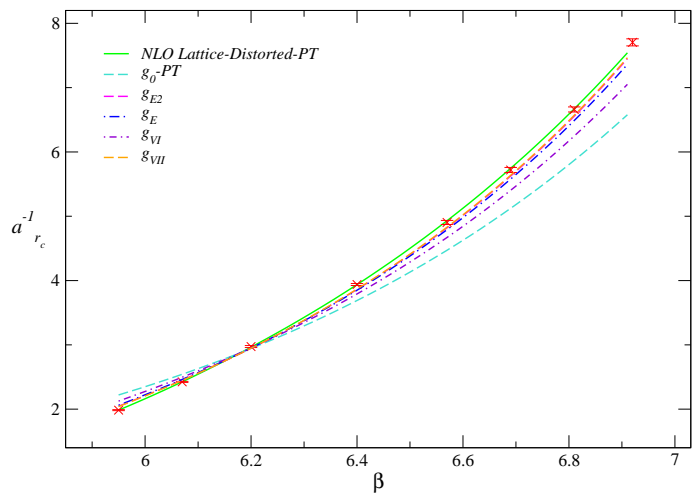}
\caption{Plot of data points $\ia$ obtained from $r_c$ with the
Wilson action together with fits from 
{\em next-to-leading order lattice-distorted} PT and from PT in different 
renormalised schemes (see text).
The data was taken from \cite{Necco}.}
\label{fig:rc+allfit}
\end{center}
\end{figure}


 
\section{Lattice-Distorted PT}
\label{sec:LDPT}

Eq.~(\ref{eq:ia_LDPT}) can be written in a more useful form
by including the leading ${\cal O}(a^n)$ and next-to-leading
${\cal O}(a^{n'})$ terms only:
:
\begin{equation}
\ia_L(g^2_0)= \frac{\Lambda_L}{f_{PT}(g^2_0)}
\times \left[ 1 
- \x\frac{g_0^{\nu} f^n_{PT}   (g^2_0)}{G_0^{\nu} f^n_{PT}   (G^2_0)}
- \y\frac{g_0^{\nu'}f^{n'}_{PT}(g^2_0)}{G_0^{\nu'}f^{n'}_{PT}(G^2_0)}
\right],
\label{eq:ia_NLO}
\end{equation}
where $G_0$ is some convenient, reference value of $g_0$, e.g. we
take $G_0 = 1$ for the Wilson data (corresponding to $\beta=6$).
In Eq.~(\ref{eq:ia_LDPT}), the values of $n, n', \nu$ and
$\nu'$\footnote{Normally $\nu \equiv \nu'$.}
depend on the lattice action being used in the Monte Carlo simulations,
and on the quantity being used to set the scale, $a$ (see Table~1).
Eg. for $T_c$ with the Wilson action, we have $n = 2$,
$n'=4$, and $\nu = \nu' = 0$.

The physical quantities used to set $a$ in this study are:
the string tension $\sigma$ \cite{Bali&Sch,CP-PACS};
the length scale $r_0$  and $r_c$ \cite{Necco,CP-PACS},
and the critical temperature $T_c$ \cite{Necco}.
The $K, K^*$ mass point on the plot of
$M_V$ versus $M^2_{PS}$, where $M_V$ and $M_{PS}$ are the mass of 
the vector and pseudoscalar meson respectively \cite{leonardo},
provides the fifth source of $\ia$ data \cite{CP-PACS}.
As well as the Wilson action, data from the Iwasaki and DBW2
action were also studied.


\begin{table}
\begin{center}
\begin{tabular}{|lcc|lllc|}
\hline
\hline
\multicolumn{1}{|c}{Action} & \multicolumn{1}{c}{Data taken}
& \multicolumn{1}{c}{Lattice data}
& \multicolumn{1}{|c}{$\Lambda_L$[Mev]} & \multicolumn{1}{c}{$\x$}
& \multicolumn{1}{c}{$\y$}
& \multicolumn{1}{c|}{$\chi^2$/\em dof}\\
 & from &&&&& \\
\hline
\hline
       &\cite{Necco}   &$r_c$   & 7.54(4)&.23(2)&-.03(2)&.69  \\
Wilson &\cite{Necco}   &$T_c$   & 7.53(3)&.213(5)&-.016(2)& .87 \\
       &\cite{Bali&Sch}&$\sigma$& 7.02(9)&.25(2)&-.02(1)&.19 \\
\hline
       &\cite{Necco,CP-PACS}&$r_0$&    3.62(2)&.046(1)&-.0106(6)&.15\\
Iwasaki&\cite{Necco}        &$T_c$&    3.68(2)&.121(3)&-.024(1)&14  \\
       &\cite{CP-PACS}      &$\sigma$& 3.49(4)&.13(1)&-.028(4)&.27\\
       &\cite{CP-PACS}      &$K-K^*$&  4.6(5)&.22(6)&-.07(3)&.60 \\
\hline
DBW2   &\cite{Necco} &$r_0$ &1.24(3) &.064(6) &-.019(4)& 4.1   \\
       &\cite{Necco} &$T_c$ &1.248(4)&.2756(6)&$-$ & 3.7        \\ 
\hline
\hline
\end{tabular}
\end{center}
\caption{Results from the fits of lattice data to {\em lattice distorted PT},
using the 3-loop $f_{PT}$ for the Wilson action and the 2-loop $f_{PT}$ for
the Iwasaki and DBW2 actions .}
\end{table}



\begin{figure}
\begin{center}
\includegraphics[angle=0,width=0.7\textwidth]{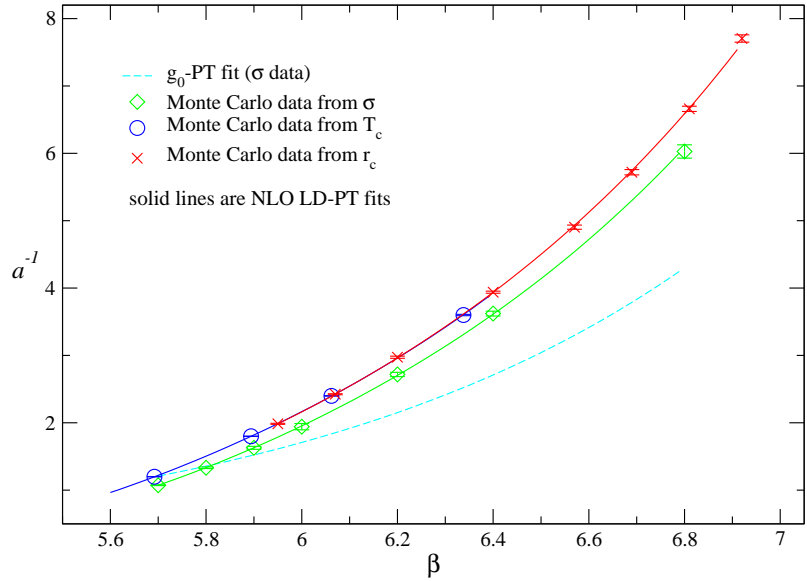}
\caption{Plot of the Monte Carlo data $\ia$ obtained from observables with 
the Wilson action together with the {\em NLO lattice-distorted} PT curves. The 
3-loop $f_{PT}$ function was used.}
\label{fig:wilson_inva}
\end{center}
\end{figure}



\begin{figure}
\begin{center}
\includegraphics[angle=0,width=0.7\textwidth]{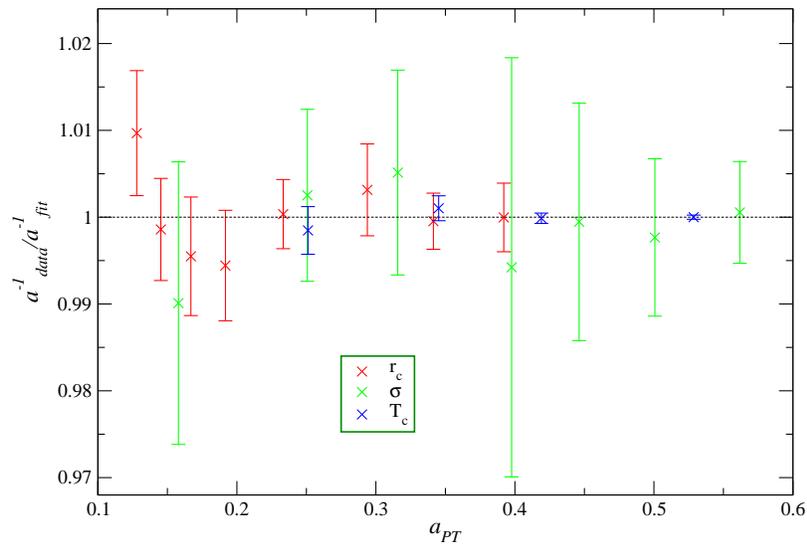}
\caption{Plot of $\ia_{data}/\ia_{fit}$ versus $\ia_{PT}$ for all data
in the Wilson case, where $\ia_{PT} = \Lambda_L/f_{PT}(g^2_0)$. Data from
the NLO fit and the 3-loop $f_{PT}$ function were used. }
\label{fig:a_pt_wilson}
\end{center}
\end{figure} 


Where possible we fit data to
next-to-leading-order (i.e. Eq.~(\ref{eq:ia_NLO})) obtaining
the coefficients $\x$ and $\y$ shown in Table 1. At NNLO the results
generally don't change from NLO, so we consider the results in Table 1  as our
best fits.
The only exception to this is the case of $K, K^*$  mass point,
where the NNLO fit finds a $\Lambda_L$ value much closer to that
from other quantities.
The DBW2 fit for $T_c$ was constrained to be at LO due to the small number of
data points available.
The Wilson data was fitted to 3-loops, whereas the Iwasaki and DBW2 actions
to 2-loops.

\begin{figure}
\begin{center}
\includegraphics[angle=0,width=0.7\textwidth]{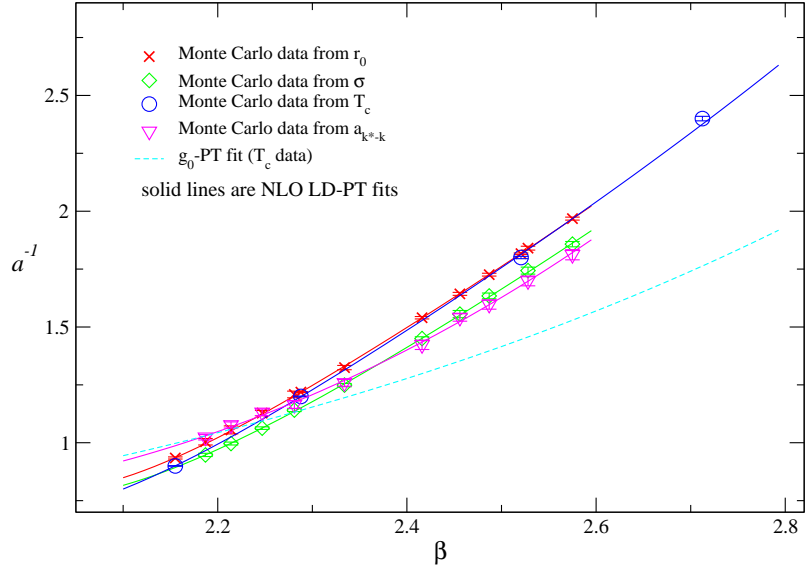}
\caption{Plot of the Monte Carlo data $\ia$ obtained from observables with 
the Iwasaki action together with the {\em NLO lattice-distorted} PT curves.
 The 2-loop $f_{PT}$ function was used.}
\label{fig:iwasaki_inva}
\end{center}
\end{figure}

\begin{figure}
\begin{center}
\includegraphics[angle=0,width=0.7\textwidth]{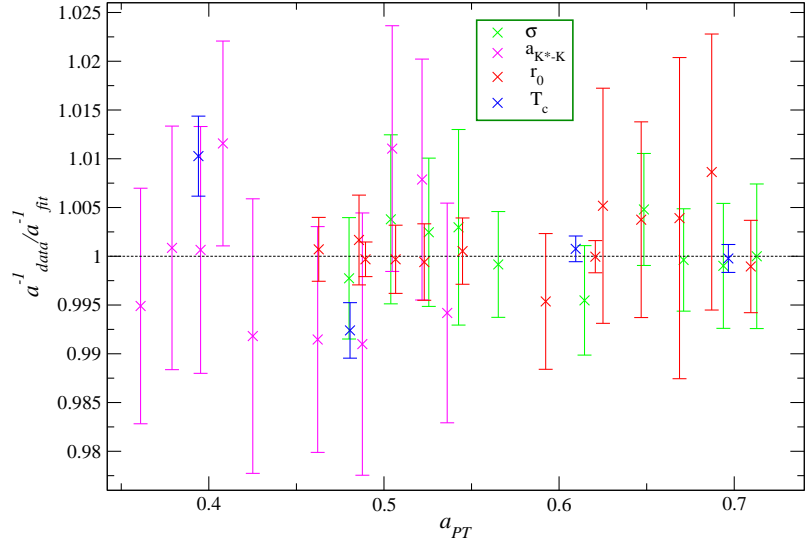}
\caption{Plot of $\ia_{data}/\ia_{fit}$ versus $\ia_{PT}$ for all data
in the Iwasaki case, where $\ia_{PT} = \Lambda_L/f_{PT}(g^2_0)$. Data from
the NLO fit and the 2-loop $f_{PT}$ function were used.}
\label{fig:a_pt_iwasaki}
\end{center}
\end{figure}

We plot the data and fits for $a^{-1}$ versus $\beta$
in Figures \ref{fig:wilson_inva} and \ref{fig:iwasaki_inva}
for the Wilson and Iwasaki actions respectively.
In order to show the high level of agreement between the
data and lattice distorted perturbation theory, we plot
the ratio of the data to the fit in 
\ref{fig:a_pt_wilson} and \ref{fig:a_pt_iwasaki} showing agreement
at the percent level.


\section{Conclusions}

Note that the quality of the fits (as indicated by the $\chi^2$ in
Table 1, and Figs. 3 \& 5) is generally excellent.
Furthermore, the $\Lambda_L$ value for each action is remarkably
consistent (see remarks in Sec.~\ref{sec:LDPT} regarding the
$K-K^*$ fit).
Note also that the series in $a$ appears convergent, in that
$\y << \x$ for all cases.
Finally, $\x$, which is a measure of the leading-order lattice systematics,
is $\sim 20$\% in the Wilson case, and, as expected,
it is significantly smaller for the improved actions.
(Note that there are only three $T_c$ data points for the DBW2 action,
hence the fit parameters in this case can be discounted.)

Converting $\Lambda_L$ for the $r_c$ and $T_c$ Wilson case to
$\Lambda_{\overline{MS}}$ \cite{dashen}, we obtain\footnote
{We are still investigating the extraction
of $\Lambda_{\overline{MS}}$ from the Iwasaki and DBW2 fits.}
\be
\Lambda^{N_f=0}_{\overline{MS}} = 217\pm 21\;\;Mev ,
\label{eq:Lambda_MS}
\ee
in agreement with previous lattice determinations in quenched QCD
\cite{Chris,Bali&Sch}.

Finally, since the lack of perturbative scaling is probably due to a mixture 
of both lattice artefacts and the poor convergence of the $g_0-$PT,
we have started to fit data to Lattice-Distorted PT
using a renormalized coupling constant instead of $g_0$
(see \cite{bothresults}).
In particular, in the $g_E$ scheme where the 3-loop coefficient is known, we 
obtain results consistent with \cite{Bali&Sch}, but the addition of the
${\cal O}(a^n)$ terms improves the quality of the fit.





\begin{thebibliography}{99}

\bibitem{Necco}
S.~Necco,
{\em Ph.D. Thesis (Humboldt U., Berlin) June 2003,}
{\tt [arXiv:hep-lat/0306005]}.


\bibitem{Lep&Mack}
G.P.~Lepage and P.B.~Mackenzie,
Phys.\ Rev.\ {\bf D48} (1993) 2250,
{\tt [arXiv:hep-lat/9209022]}.

\bibitem{Parisi}
G.~Parisi,
LNF-80/52-P
{\it Presented at 20th Int. Conf. on High Energy Physics, Madison, Wis., Jul 17-23, 1980}

\bibitem{Bali&Sch}
G.S.~Bali and K.~Schilling,
Phys. Rev. {\bf D47} (1993) 661,
{\tt [arXiv:hep-lat/9208028]}.

\bibitem{Chris}
C.R.~Allton,
{\tt [arXiv:hep-lat/9610016]};
Nucl.\ Phys.\ {\bf B(Proc.Suppl.) 53} (1997) 867,
{\tt [arXiv:hep-lat/9610014]}

\bibitem{CP-PACS}
A.~Ali Khan {\it et al.}  [CP-PACS Collaboration],
Phys.\ Rev.\ {\bf D65} (2002) 054505,
[Erratum-ibid.\ D {\bf 67} (2003) 059901],
{\tt [arXiv:hep-lat/0105015]}.

\bibitem{leonardo}
C.R.~Allton, V.~Gim\`enez, L.~Giusti, and F.~Rapuano, \\
Nucl.\ Phys.\ {\bf B489} (1997) 427
{\tt [arXiv:hep-lat/9611021]}

\bibitem{dashen}
R.~Dashen and D.J.~Gross,
Phys. Rev. {\bf D23} (1981) 2340.

\bibitem{bothresults}
R.G.~Edwards, U.M.~Heller, T.R.~Klassen,
Nucl.\ Phys.\ {\bf B517} (1998) 377,
{\tt [arXiv:hep-lat/9711003]}.

                                                                               
\end{thebibliography}
\end{document}